\documentclass{JHEP3}

\usepackage{url,amsmath,amsthm,amscd,graphicx,psfrag}
\usepackage[all]{xy}

\def\tX{{\tilde X}}
\def\tV{{\tilde V}}
\def\pt{{\rm pt}}

\let\a=\alpha   \let\b=\beta        
              
     \let\m=\mu      
      \let\x=\xi                  
\let\t=\tau

 \def\cf{{\cal F}}  
   
\def\cm{{\cal M}}  \def\co{{\cal O}}

 \def\IC{{\mathbb C}} \def\IP{{\mathbb P}}
   
\def\IZ{{\mathbb Z}}

\theoremstyle{definition}

\theoremstyle{plain}

\theoremstyle{remark}

\newtheorem*{ackn}{Acknowledgments}

\def\plb#1 #2 {Phys. Lett. {\bf B#1} #2 }
\def\phr#1 #2 {Phys. Rep. {\bf  #1} #2 }    
\def\npb#1 #2 {Nucl. Phys. {\bf B#1} #2 }
\def\aph#1 #2 {Ann. Phys. {\bf #1} #2 }      
\def\jmp#1 #2 {J. Math. Phys. {\bf #1} #2 }
\def\jgp#1 #2 {J. Geom. Phys. {\bf #1} #2 }
\def\prd#1 #2 {Phys. Rev. {\bf D#1} #2 }
\def\prl#1 #2 {Phys. Rev. Lett. {\bf #1} #2 }
\def\rmp#1 #2 {Rev. Mod. Phys.  {\bf #1} #2 }
\def\zpc#1 {Z. Phys. {\bf #1C} }
\def\cmp#1 #2 {Commun. Math. Phys. {\bf #1} #2 }
\def\cqg#1 #2 {Class.Quant.Grav. {\bf #1} #2 }
\def\mpl#1 {Mod. Phys. Lett. {\bf A#1} }
\def\cpc#1 {Computer Phys. Commun. {\bf #1} }   
\def\ijmp#1 {Int. J. Mod. Phys. {\bf A#1} }
\def\ijmpC#1 {Int. J. Mod. Phys. {\bf C#1} }
\def\atmp#1 {Adv. Theor. Math. Phys. {\bf #1} }

\numberwithin{equation}{section}

\let\0=\over
\def\2{{1\over2}}
\let\6=\partial
\def\({\left(}       \def\){\right)}

\let\bra=\langle        \let\ket=\rangle        \def\<#1\>{\bra #1 \ket}
   
\title{An $SU(5)$ Heterotic Standard Model}

\author{Vincent Bouchard\\
Department of Mathematics\\
University of Pennsylvania\\
Philadelphia, PA 19104-6395\\
E-mail: \email{vincentb@math.upenn.edu}}

\author{Ron Donagi\\
Department of Mathematics\\
University of Pennsylvania\\
Philadelphia, PA 19104-6395\\
E-mail: \email{donagi@math.upenn.edu}}

\abstract{We introduce a new heterotic Standard Model which has precisely the spectrum of the Minimal Supersymmetric Standard Model (MSSM), with no exotic matter. The observable sector has gauge group $SU(3)_C \times SU(2)_L \times U(1)_Y$. Our model is obtained from a compactification of heterotic strings on a Calabi-Yau threefold with $\IZ_2$ fundamental group, coupled with an invariant $SU(5)$ bundle. Depending on the region of moduli space in which the model lies, we obtain a spectrum consisting of the three generations of the Standard Model, augmented by $0$, $1$ or $2$ Higgs doublet conjugate pairs. In particular, we get the first compactification involving a heterotic string vacuum (i.e. a {\it stable} bundle) yielding precisely the MSSM with a single pair of Higgs.}

\begin{document}

\section{Introduction}

In this note we present the first heterotic M-theory vacua with exactly the minimal supersymmetric Standard Model spectrum and a single Higgs pair. We do not think that this compactification is ``the model of the Universe": we expect that in a few years many such models will be known. But for now, this is the best we have. :-)

A heterotic vacuum is specified by a solution of the Hermitian Yangs-Mills equation on a Calabi-Yau threefold $X$. Geometrically, this solution is determined by a stable holomorphic $G$-bundle $V$ on $X$, for an appropriate structure group $G \subset E_8$ (or sometimes $G \subset E_8 \times E_8$). 

In order to obtain the Standard Model, we take $X$ to be non-simply connected, with a Calabi-Yau cover $\tX \to X$. The stable bundle $V$, which gives a GUT gauge group, can then be twisted by a non-trivial Wilson line on $X$ which breaks the gauge group to the Standard Model group $SU(3)_C \times SU(2)_L \times U(1)_Y$. The bundle $V$ is subject to a series of constraints which assure anomaly cancellation, three generations of quarks and leptons, no exotic particles, no dangerous interactions, etc.

Recent progress towards construction of heterotic vacua which closely resemble the observed Standard Model world has been based on the application of powerful algebro-geometric techniques. These combine the spectral cover construction of stable bundles on elliptic fibrations \cite{Friedman:1997yq,Donagi:1997pb} with extension techniques.

In \cite{Donagi:2000si,Donagi:2000sm,Donagi:2000zf}, a family of Standard Model stable $SU(5)$ bundles on a Calabi-Yau $X$ with fundamental group $\pi_1 (X) = \IZ_2$ were constructed. The spectrum was determined in \cite{Donagi:2004su,Donagi:2004ub}; in addition to the three generations of quarks and leptons it contains various exotic particles.

Stable $SU(4)$ bundles on a Calabi-Yau threefold with $\pi_1 (X) = \IZ_2 \times \IZ_2$ were constructed in \cite{Donagi:2003tb}. While interesting mathematically, these do not lead to the Standard Model gauge group since the latter cannot be obtained as a centralizer of $\IZ_2 \times \IZ_2$ in the GUT gauge group $Spin(10)$.

Very recently, in an interesting series of papers \cite{Braun:2004xv,Braun:2005ux,Braun:2005bw,Braun:2005zv} a family of $SU(4)$ bundles on a Calabi-Yau $X$ with fundamental group $\pi_1 (X) = \IZ_3 \times \IZ_3$ was constructed using extension techniques. Although it has been shown that these bundles pass various highly non-trivial stability checks, it is still unknown whether these bundles are stable. So it is not clear whether there are any corresponding physical vacua. If they turn out to be stable, which seems probable, they would yield the Standard Model spectrum plus two pairs of Higgs doublets. The Yukawa couplings for these bundles vanish identically. 

In this note we describe a new family of heterotic M-theory vacua given by {\it stable} $SU(5)$ bundles on the $\IZ_2$ manifolds of \cite{Donagi:2000si}. The moduli space of the family has many regions. The spectrum in three of those regions consists precisely of the three generations of the Standard Model, augmented by $0$, $1$ or $2$ pairs of Higgs doublets. As one moves from one region of the moduli space to the other, the number of Higgs pairs jumps --- the possibility of such a jump was first noted in \cite{Donagi:2004qk,Donagi:2004ia}. In particular, the middle region of our family has exactly the expected spectrum.\footnote{There are also additional regions in the moduli space of our model where Higgs doublets are accompanied by Higgs triplets, which are exotic.}

The construction is surpisingly close to that of \cite{Donagi:2000sm,Donagi:2000zf,Donagi:2004ub}: the $SU(5)$ bundle $V$ is constructed as an extension involving bundles $V_2'$ and $V_3'$ of rank $2$ and $3$ respectively. The main change is that for our new bundle, $V_3'$ is the subbundle and $V_2'$ the quotient; in \cite{Donagi:2000sm} this was reversed. Equivalently, we can describe the dual: a rank $5$ bundle with a subbundle $V_2$ and a quotient $V_3$, but with $3$ ``anti-generations" instead of generations.

Recently the authors of \cite{Braun:2005ux} informed us that they have constructed a new family of bundles on the manifold of \cite{Braun:2005ux} which yields only one pair of Higgs doublets \cite{Braun:un}. But again it is unclear whether their new bundle is stable or not.

\subsection{Main Result}

The observable sector of our compactification of the $E_8 \times E_8$ heterotic string has the following properties:
\begin{itemize}
\item Gauge group: $SU(3)_C \times SU(2)_L \times U(1)_Y$;
\item Three families of quarks and leptons;
\item $0$, $1$ or $2$ Higgs doublet conjugate pairs;
\item A certain number of moduli;
\item No exotic matter fields.
\end{itemize}

Finally, the hidden sector has gauge group $E_8$ and contains no matter fields or moduli.

\subsection{Outline}

Since our construction is very similar to the construction of \cite{Donagi:2000si,Donagi:2000sm,Donagi:2000zf,Donagi:2004ub}, we tried to keep this note brief and only summarize the results of these papers. The details can be found in the above references.

In section \ref{s:man} we briefly describe the Calabi-Yau threefold used in the compactification. We construct the bundle in section \ref{s:bundle}, and show that it is $\t$-invariant, stable, and that it satisfies Standard Model constraints. We compute the relevant cohomology groups and study their invariant and anti-invariant parts in section \ref{s:coh}. Finally, we describe the particle spectrum in section \ref{s:spectrum}.

\begin{ackn}
The authors would like to thank Volker Braun, Burt Ovrut and Tony Pantev for interesting discussions on this subject. V.B. would like to thank the Department of Mathematics and the Department of Physics of University of Pennsylvania for hospitality during the completion of this work. R.D. is partially supported by an NSF grant DMS 0104354. Both authors are partially supported by an NSF Focused Research Grant DMS 0139799 for ``The Geometry of Superstrings".
\end{ackn}

\section{The Manifold}\label{s:man}

The Calabi-Yau threefold is $X$ constructed by considering a simply connected Calabi-Yau threefold $\tilde X$, elliptically fibered over a rational elliptic surface, that admits a free $F=\IZ_2$ action (an ``involution") preserving the fibration (but not the section). The quotient $X = {\tilde X} / F$ is a Calabi-Yau threefold, has fundamental group $F$ and is a genus-one fibration.

The Calabi-Yau threefold $\tX$ that we consider here was constructed in \cite{Donagi:2000si,Donagi:2004ub}. Therefore we will only sketch the construction; the reader is referred to the above papers for a detailed analysis of its properties.

Let $B'$ be a rational elliptic surface, and let $\tilde X$ be a Calabi-Yau threefold with an elliptic fibration $\pi: {\tilde X} \to B'$ (we also require that $\pi$ has a section). The manifold $\tilde X$ also admits a description as a fiber product $B \times_{\IP^1} B'$ of two rational elliptic surfaces $B$ and $B'$ over $\IP^1$:
\begin{equation}
{\tilde X} = \{ (p,p') \in B \times B' | \b'(p') = \b (p) \},
\end{equation}
 where $\b: B \to \IP^1$ and $\b': B' \to \IP^1$ are the elliptic fibrations of the rational elliptic surfaces $B$ and $B'$.

Thus $\tilde X$ can be described by the following commuting diagram
\begin{equation}\label{e:fibration}
\xymatrix{
& {\tilde X} \ar[dl]_{\pi'} \ar[dr]^{\pi} \\
B \ar[dr]^{\b} && B' \ar[dl]_{\b'} \\
& \IP^1
}
\end{equation}

The two rational elliptic surfaces $B$ and $B'$ are chosen such that they lie in the four-parameter family of rational elliptic surfaces described in \cite{Donagi:2000si,Donagi:2004ub}. Both of them admit a $\IZ_2$ involution $\t_B$ and $\t_{B'}$ respectively. The involutions $\t_B$ ($\t_{B'}$) split into an involution that preserves the zero section $\a_B$ ($\a_{B'}$) composed by a translation by a section $t_\x$ ($t_{\x'}$). $\t_B$ has only four fixed points in the fiber $f_{\infty}$ above $\infty \in \IP^1$, while $\t_{B'}$ has only four fixed points in the fiber $f_0$ above $0 \in \IP^1$. Thus, if we identify the $\IP^1$s in the two elliptic fibrations $\b$ and $\b'$ we find a fiber product $\tX= B \times_{\IP^1} B'$ that admits a free $\IZ_2$ involution $\t := \t_B \times_{\IP^1} \t_{B'}$.

A rational elliptic surface\footnote{Indeed, all the following facts hold for $B'$ as well.} $B$ is the blow up of $\IP^2$ in nine points. The cohomology group $H^2(B, \IZ) = Pic(B)$ has rank $10$, and we can choose as an orthogonal basis the class $\ell$, which is the pullback of the hyperplane class of $\IP^2$, and the nine exceptional classes $e_1, \ldots, e_9$. The only non-zero intersection numbers are given by $\ell^2 = 1$, $e_i^2=-1$, $i=1,\ldots,9$. The canonical class of $B$ is given by $K_B = -f$, where $f$ is the fiber class of $\b$ and can be expressed in the above basis by $f = 3 \ell - \sum_i e_i$. In fact, the exceptional divisors $e_i$, $i=1,\ldots,9$ are sections of the elliptic fibration $\b$; we define $e_9$ to be the zero section, and $\x:= e_1$ to be the section used to define the involution $t_\x$.

\section{The Bundle}\label{s:bundle}

The bundle we consider is very similar to the ones studied in \cite{Donagi:2000si,Donagi:2000sm,Donagi:2004ub}, where a systematic analysis of this kind of bundle was done. However, in their analysis a few cases were missed; it turns out that one of them gives precisely the spectrum of the MSSM.

As usual, to get an $SU(5)$ bundle $V$ on $X$, we construct an $SU(5)$ bundle $\tV$ on $\tX$ together with an action of the involution $\t$ on $\tV$.

Instead of working directly with the bundle $\tV$, in the following we will consider its dual $\tV^*$, since in that case we can apply directly the results of \cite{Donagi:2000sm,Donagi:2004ub}. The bundle $\tV^*$ is constructed as an extension
\begin{equation}\label{e:bdef}
0 \to V_2 \to \tV^* \to V_3 \to 0,
\end{equation}
where $V_2$ and $V_3$ are rank $2$ and $3$ bundles respectively\footnote{Notice that $\tV$ is then given by an extension with a rank $3$ subbundle $V'_3$ and a rank $2$ quotient $V'_2$, rather than the opposite; this is precisely why this case was missed in the analysis of \cite{Donagi:2000sm,Donagi:2004ub}}.

The bundles $V_i$ are given by
\begin{equation}
V_i = \pi'^* W_i \otimes \pi^* L_i,
\end{equation}
where the $L_i$ are some line bundles on $B'$ and the $W_i$ are rank $i$ bundles on $B$ given by the Fourier-Mukai transforms $W_i = FM_B (C_i,N_i)$: as usual in the spectral cover construction, the $C_i \subset B$ are curves in $B$ and the $N_i \in Pic(C_i)$ are line bundles over $C_i$.

In order to satisfy all the Standard Model constraints, we choose the following data:
\begin{align}
{\bar C_2} &\in | \co_B ( 2 e_9 +2f)|,\notag \\
C_3 &\in | \co_B ( 3 e_9 + 3 f)|,\notag\\
C_2 &= {\bar C_2} + f_{\infty},\notag\\
N_2 &\in Pic^{3,1} (C_2),\notag\\
N_3 &\in Pic^7 (C_3),\notag\\
L_2 &= \co_B' ( 3 r'),\notag\\
L_3 &= \co_B' ( -2 r').
\end{align}
$f_{\infty}$ is the smooth fiber of $\b$ at $\infty$ containing the four fixed points of $\t_B$, and $Pic^{3,1} (C_2)$ denotes line bundles of degree $3$ over $\bar C_2$ and degree $1$ on $f_{\infty}$. Finally, $r'$ is given by
\begin{equation}
r' = e_1' + e_4' -e_5' + e_9' + f'.
\end{equation}

It is easy to see that our $V_2$ is exactly the same as the one considered in \cite{Donagi:2004ub}; the only difference between our bundle and their bundle is in the spectral data of $W_3$. Therefore most of their results will be valid for our construction as well, although the minor modifications will turn out to be crucial.

\subsection{Invariance}

The first condition is that the bundle $\tV^*$ must be $\t$-invariant. It is shown in section $4$ of \cite{Donagi:2000sm} that if the $V_i$ are $\t$-invariant and the space parametrizing all extensions ${\rm Ext}^1 ( V_3, V_2 ) = H^1 (\tX, V_2 \otimes V_3^*)$ is non-zero, then there are non-trivial extensions involving $V_2$ and $V_3$ which are also $\t$-invariant. Thus we will require that $V_2$ and $V_3$ be $\t$-invariant.

It was shown in \cite{Donagi:2000sm,Donagi:2004ub} that $V_2$ is $\t$-invariant. Thus we only have to show that $V_3$ is $\t$-invariant. Recall that $L_3 = \co(-2r')$; $r'$ is an invariant homology class on $B'$, therefore $L_3$ is $\t$-invariant. All that is left is to check that $W_3$ is also invariant.

As explained in \cite{Donagi:2000sm}, to check that $W_3$ is invariant we must verify that the spectral data satisfies
\begin{align}
C_3 &= \a_B (C_3),\notag\\
N_3 & = {\bf T}_{C_3} (N_3),
\end{align}
where $\a_B$ is the involution on $B$ that fixes the zero section and ${\bf T}_{C_3}$ is the spectral involution induced by $\t$.

First, lemma 4.1 of \cite{Donagi:2000sm} tells us that the linear system $|\co(3e_9 + 3f)|$ contains smooth $\a_B$-invariant curves. Therefore, we choose such a $C_3 \in |\co(3e_9 + 3f)|$, and the first condition is satisfied.

Now, lemma 4.3 of \cite{Donagi:2000sm} tells us that for every $d \in \IZ$ there exists line bundles $N_3 \in Pic^d (C_3)$ such that $N_3={\bf T}_{C_3} (N_3)$, if $i_{C_3}^* Pic (B)$ is dense in $Pic^0 (C_3)$. Thus we have to check that the last condition is satisfied. As the proof is rather technical, it is presented in the Appendix. The result is that for $C_3 \in |\co(3e_9 + 3f)|$ an $\a_B$-invariant curve, $i_{C_3}^* Pic (B)$ is dense in $Pic^0 (C_3)$, and thus we can choose an $N_3 \in Pic^7 (C_3)$ which is ${\bf T}_{C_3}$-invariant, as required.

\subsection{Standard Model Constraints}

To have a Standard Model bundle, three conditions on the Chern classes must be satisfied. First, $c_1 (\tV^*)=0$ for the bundle to have structure group $SU(5)$ rather than $U(5)$; second, anomaly cancellation in the heterotic string imposes that $c_2(X)-c_2(\tV^*)$ be an effective class (around which $M5$-branes wrap to cancel the anomaly); third, $c_3 (\tV^*) = \pm 12$ to obtain three generations of particles.

It is easy to compute the Chern character of our bundle. First, we have
\begin{align}
ch (W_2) &= 2-f-3 \pt,\notag\\
ch (W_3) &= 3+f-3 \pt.
\end{align}
Combining with the Chern character of the line bundles, and from the fact that the Chern character of $\tV^*$ is the sum of the Chern characters of $V_2$ and $V_3$, we obtain
\begin{equation}
ch (\tV^*) = 5 - 10 (f \times \pt) - 6 ( \pt \times f') -6 \pt.
\end{equation}
Thus $c_1 (\tV^*) = 0$, $c_3 (\tV^*) = -12$, and we find
\begin{align}\label{e:c2}
c_2(X)-c_2(\tV^*) &= 12(f \times \pt + \pt \times f') - 10 (f \times \pt) - 6 ( \pt \times f')\notag\\
&= 2 f \times \pt + 6 \pt \times f',
\end{align}
which is effective. A consequence of \eqref{e:c2} is that we do not need to add a gauge instanton in the hidden sector to cancel the anomaly. It also tells us that we are in the strong coupling regime of the heterotic string, with $M5$-branes wrapping the effective curve given by $2 f \times \pt + 6 \pt \times f'$. It may be possible to add a gauge instanton $U$ of small rank in the hidden sector such that $c_2 (U) = 2 f \times \pt + 6 \pt \times f'$, which would give a weak coupling vacuum of our model; we have not considered this possibility yet.

The reason why this model was missed in \cite{Donagi:2000sm} is that $c_3 (\tV^*) = -12$ rather than $c_3 (\tV^*) = 12$. As we mentioned already, this implies that the dual $\tV$ will be an extension with rank $3$ subbundle and a rank $2$ quotient bundle rather than the opposite.

\subsection{Stability}

A necessary condition for the bundle to provide a well defined compactification of heterotic string theory is that it is stable with respect to a fixed ample line bundle (or K\"ahler structure) $A$ on $\tX$.

It was shown in lemma 5.3 of \cite{Donagi:2000sm} that the extension $\tV^*$ defined as in the previous section is $A$-stable if and only if the extension $\tV^*$ is non-trivial and the slope of $V_2$ satisfies $\m_A (V_2) < 0$.

If $A_0$ is any K\"ahler class on $\tX$, $h'$ a K\"ahler class on $B'$, and $n \gg 0$, then the class $A = A_0 + n \pi^* h'$ is K\"ahler on $\tX$. As in \cite{Donagi:2004ub}, we choose $h' = 193 f' + 144 e_1' + 168 ( e_9' + e_4' - e_5')$. For this value it was shown in \cite{Donagi:2000sm} that $\m_A (V_2) < 0$.

So we only have to make sure that there exists a non-trivial extension involving $V_2$ and $V_3$, that is ${\rm Ext}^1 ( V_3, V_2 ) = H^1 (\tX, V_2 \otimes V_3^*) \neq 0$. Lemma 5.4 of \cite{Donagi:2000sm} tells us that this will be the case if $L_2 \cdot f' > L_3 \cdot f'$. But in our case, $L_2 \cdot f' = 6$ and $L_3 \cdot f' = -4$, so $H^1 (\tX, V_2 \otimes V_3^*) \neq 0$.

We have thus proven that if we choose $\tV^*$ to be a non-trivial extension involving $V_2$ and $V_3$, then it is $A$-stable. Although this seems like a mathematical detail, it is a crucial point, since the compactification of heterotic strings is simply not well defined if the bundle is unstable.

\section{The Cohomology}\label{s:coh}

In this section we compute the relevant cohomology groups. Most computations of \cite{Donagi:2004ub} are still valid with only minor modifications. Let us first recall the setup used in the computations of \cite{Donagi:2004ub}. Define a quadric surface $Q=\IP^1_x \times \IP^1_t$. The rational elliptic surface $B'$ is a double cover of $Q$, given by the map $\triangle: B' \to Q$. Define the coordinate $y$ to be a section of $\co_{B'} (2 r'+ f')$ which vanishes on the ramification locus of the map $\triangle: B' \to Q$. Denote by $S_x^k := H^0 (\co_{\IP^1_x} (k))$ the $(k+1)$-dimensional vector space of homogeneous polynomials of degree $k \geq 0$ in $x_0,x_1$, with basis consisting of the monomials $x_0^k, x_0^{k-1} x_1, \ldots, x_1^k$ ($S_x^{k*}$ is the dual vector space). $S^k_t$ is the similar space for $\IP^1_t$. Then it turns out that the relevant cohomology groups can be expressed explicitly in terms of $S^j_x$, $S^k_t$ and $y$. This point of view on the rational elliptic surface $B'$ is described in detail in \cite{Donagi:2004ub}.

\begin{list}{$\bullet$}{\setlength{\leftmargin}{0.4cm}}
\item
\boxed{V_2}
It was computed in \cite{Donagi:2004ub} that  $h^0 ( \tX, V_2) = 0$, $h^1 ( \tX, V_2 ) = 6$.
\item
\boxed{V_3}
As in \cite{Donagi:2004ub}, we find that $h^0 ( \tX, V_3) =h^1 ( \tX, V_3) = 0$.
\item
\boxed{\tV^*}
The long exact sequence in cohomology tells us that $h^1 ( \tX, \tV^*) = 6$. Since $h^0( \tX, V_2) =h^3( \tX, V_2)=h^0( \tX, V_3) =h^3( \tX, V_3)=0$, from the index theorem and the fact that $c_3 (\tV^*) = -12$ we obtain
\begin{equation}
-h^1 ( \tX, \tV^*) + h^2 ( \tX, \tV^*) = -6,
\end{equation}
thus $h^2 ( \tX, \tV^*) = 0$.
\item
\boxed{\wedge^2 V_2}
The cohomology groups were computed explicitly in \cite{Donagi:2004ub}, where it was found that
\begin{align}
H^0 (\tX, \wedge^2 V_2) &= 0, &H^1 (\tX, \wedge^2 V_2) &= y S_x^4,\notag\\
H^2 (\tX, \wedge^2 V_2) &= S_x^6 \oplus y S_x^4 \otimes S^{1*}_t, &H^3 (\tX, \wedge^2 V_2) &= 0.
\end{align}

\item
\boxed{\wedge^2 V_2^*}
These cohomology groups can be obtained from the cohomology groups of $\wedge^2 V_2$ by Serre duality. We find
\begin{align}
H^0 (\tX, \wedge^2 V_2^*) &= 0, &H^1 (\tX, \wedge^2 V_2^*) &= S_x^{4*} \otimes S_t^1 \oplus y S_x^{6*},\notag\\
H^2 (\tX, \wedge^2 V_2^*) &= S_x^{4*}, &H^3 (\tX, \wedge^2 V_2^*) &= 0.
\end{align}

\item
\boxed{V_2 \otimes V_3^*}
The analysis of \cite{Donagi:2004ub} carries through with a minor (although crucial) modification. The intersection of $f_{\infty}$ and $C_3$ still consist in three points $p_j$, $j=7,8,9$. But the curves ${\bar C_2}$ and $C_3$ now intersect in $6$ points rather than $12$ points. Therefore, the image of $\b({\bar C_2} \cap C_3)$ consists now in $6$ distinct points ${\hat p_j} \in \IP^1_t$, and we find that
\begin{equation}
H^1 ( \tX, V_2 \otimes V_3^*) = \bigoplus^{9}_{j=1} H^0 (\IP^1_t, \cf_j) \otimes [S_x^5 \oplus y S_x^3 \otimes \{ {\hat p}_j \IC \}].
\end{equation}
In this equation $\{{\hat p}_j \IC \} \subset S_t^{1*}$ denotes the lines inside the two-dimensional plane $S_t^{1*}$ consisting of all points proportional to ${\hat p}_j \in \IP^1_t$. Moreover, each $\cf_j$ is a skyscraper sheaf supported on the point ${\hat p}_j$. It is easy to see that $h^1 ( \tX, V_2 \otimes V_3^*) = 90$.

\item
\boxed{V_2^* \otimes V_3^*}
Again, the only difference is in the number of intersection points, so we find
\begin{equation}
H^2 ( \tX, V_2^* \otimes V_3^*)=\bigoplus^{9}_{j=1} H^0 (\IP^1_t, \cf_j) \otimes y S_x^{1*},
\end{equation}
which has dimension $18$.

\item
\boxed{\wedge^2 \tV^*}
As in \cite{Donagi:2004ub}, we find that $h^0 (\tX, \wedge^2 \tV^*)= h^3 (\tX, \wedge^2 \tV^*)=0$. Moreover, the cohomology group $H^1 (\tX, \wedge^2 \tV^*)$ fits into the exact sequence
\begin{equation}\label{e:es}
0 \xrightarrow{} H^1 (\tX, \wedge^2 V_2) \xrightarrow{} H^1 (\tX, \wedge^2 \tV^*) \xrightarrow{} H^1 (\tX, V_2 \otimes V_3) \xrightarrow{M^T} H^2 (\tX, \wedge^2 V_2) \xrightarrow{} \ldots
\end{equation}
The crucial point lies in the analysis of the coboundary map
\begin{equation}
M^T : H^1 (\tX, V_2 \otimes V_3) \to H^2 (\tX, \wedge^2 V_2).
\end{equation}
It is given by the cup product with the extension class of our bundle:
\begin{equation}
[\tV^*] \in H^1 ( \tX, V_2 \otimes V_3^*) = {\rm Ext}^1 (V_3, V_2),
\end{equation}
via the pairing
\begin{equation}
\cm^T : H^1 (\tX, V_2 \otimes V_3) \times H^1 ( \tX, V_2 \otimes V_3^*) \to H^2 (\tX, \wedge^2 V_2).
\end{equation}
It is easier to work with its dual:
\begin{equation}
\cm : H^1 (\tX, \wedge^2 V_2^*) \times H^1 ( \tX, V_2 \otimes V_3^*) \to H^2 (\tX, \wedge^2 V_2^* \otimes V_3^*).
\end{equation}
More explicitly, $\cm$ is the product
\begin{equation}
\cm: (S_x^{4*} \otimes S_t^1 \oplus y S_x^{6*}) \otimes \left( \bigoplus^{9}_{j=1} H^0 (\IP^1_t, \cf_j) \otimes [S_x^5 \oplus y S_x^3 \otimes \{ {\hat p}_j \IC \}] \right) \to \bigoplus^{9}_{j=1} H^0 (\IP^1_t, \cf_j) \otimes y S_x^{1*}.
\end{equation}
\end{list}
The three spaces involved have respectively dimensions $17$, $90$ and $18$. $\cm$ breaks into $9$ blocks sending a $17 \times 10$ dimensional space to a $2$-dimensional space. Each block breaks further into a $10 \times 4 \to 2$ block and a $7 \times 6 \to 2$ block, as in \cite{Donagi:2004ub}; they correspond to the products
\begin{gather}
(S_x^{4*} \otimes S_t^1 ) \otimes (S_x^3 \otimes \{ {\hat p}_j \IC \} ) \to S_x^{1*}, \notag\\
(S_x^{6*}) \otimes (S_x^5) \to S_x^{1*}.
\end{gather}
The map $M$ is given by evaluating at the extension class $[\tV^*]$ of our bundle. Therefore, we must express this class in the above basis. Let $a_{i,j}$, $i=0,\ldots,5$, $j=1,\ldots,9$ and $b_{k,j}$, $k=0,\ldots,3$, $j=1, \ldots, 9$ be the coefficients of $[\tV^*]$ in the $S_x^5$ and $S_x^3$ factors respectively. Using this explicit description of the extension class, we can express the maps in matrix form. The map $S_x^{6*} \to S_x^{1*}$ is represented by the $2 \times 7$ matrix in the $a_{i,j}$ coefficients:
\begin{equation}\label{e:ma}
M_{I,j}=
\begin{pmatrix}
a_{0,j} & \ldots & a_{5,j} & 0 \\
0 & a_{0,j} & \ldots & a_{5,j}
\end{pmatrix},
\end{equation}
while the map $S_x^{4*} \otimes S_t^1 \to S_x^{1*}$ is given by the $2 \times 10$ matrix
\begin{equation}\label{e:mb}
M_{II,j} =
\begin{pmatrix}
b_{0,j} t_0 ({\hat p}_j ) & \ldots & b_{3,j} t_0 ({\hat p}_j ) & 0 &b_{0,j} t_1 ({\hat p}_j ) & \ldots & b_{3,j} t_1 ({\hat p}_j ) & 0\\
0 & b_{0,j} t_0 ({\hat p}_j ) & \ldots & b_{3,j} t_0 ({\hat p}_j ) & 0 &b_{0,j} t_1 ({\hat p}_j ) & \ldots & b_{3,j} t_1 ({\hat p}_j )
\end{pmatrix}.
\end{equation}
Thus the full $18 \times 17$ matrix is given by
\begin{equation}
M = 
\begin{pmatrix}
M_{I,1} & M_{II,1}\\
\vdots & \vdots\\
M_{I,9} & M_{II,9}
\end{pmatrix}.
\end{equation}
Now the next step is to find the rank of the matrix $M$: this will give us the dimension of the cohomology group $H^1 (\tX, \wedge^2 \tV^*)$ from the exact sequence \eqref{e:es}. For instance, if $M$ has maximal rank, that is $17$, then we find $h^1 (\tX, \wedge^2 \tV^*) = 5 + 18 - 17 = 6$, and by the index theorem $h^2 (\tX, \wedge^2 \tV^*) = 0$. However, if $M$ has rank less than $17$, say $17-n$ for an integer $n$, then we would find $h^1(\tX, \wedge^2 \tV^*) = 6+n$ and $h^2(\tX, \wedge^2 \tV^*)=n $. Thus the rank of $M$, which amounts to specifying the extension class $[\tV^* ]$, is crucial in the determination of the particle spectrum.

We mentioned earlier that for our bundle $\tV^*$ to be $\t$-invariant, we choose an invariant extension class $[\tV^*]$. This puts some constraints on the coefficients $a_{i,j}$ and $b_{k,j}$, which we must understand in order to compute the rank of $M$. Let us now analyze the action of the involution $\t$ on our bundles.

\subsection{$\IZ_2$ action}

It was shown in \cite{Donagi:2004ub} that the involution $\t$ acts on the coordinates $\{ t_0, t_1, x_0, x_1, y \}$ by
\begin{equation}
t_0 \mapsto t_0,~~~t_1 \mapsto - t_1,~~~x_0 \mapsto x_0,~~~x_1 \mapsto -x_1,~~~ y \mapsto y.
\end{equation}
Using this action we can find the induced action of $\t$ on the cohomology groups using their explicit description in terms of $S_x^k$, $S_t^j$ and $y$. For example, $S_x^4$ splits into a three-dimensional invariant subspace and a two-dimensional anti-invariant subspace. We will denote the invariant subspace by a $+$ subscript and the anti-invariant subspace by a $-$ subscript.

First, as in \cite{Donagi:2004ub} we find that the six-dimensional space $H^1 (\tX, \tV^*)$ splits into two three-dimensional subspaces, that is
\begin{equation}
h^1 (\tX, \tV^*)_+ = 3,~~~~h^1 (\tX, \tV^*)_- = 3.
\end{equation}

Now we want to use this action to understand the rank of the matrix $M$, and therefore the cohomology groups $H^1(\tX, \wedge^2 \tV^*)$ and $H^2(\tX, \wedge^2 \tV^*)$. Let us first analyze the extension class $[\tV^*]$. 

Recall that it is given by the coefficients $a_{i,j}$, $i=0,\ldots,5$, $j=1,\ldots,9$ and $b_{k,j}$, $k=0,\ldots,3$, $j=1, \ldots, 9$ in the $S_x^5$ and $S_x^3$ factors respectively. The different $j$ correspond to points ${\hat p}_j \in \IP^1_t$. The last three points, ${\hat p_j}$, $j=7,8,9$, are simply $\infty \in \IP^1_t$, as explained in \cite{Donagi:2004ub}. The other six points correspond to the images under the projection $\b$ of the six intersection points of ${\bar C}_2$ and $C_3$. Since these six points are in $\IP^1_t$ and are not $0$ or $\infty$, they must be interchanged in pairs by the involution. Thus the action of $\t$ on the extension class must interchange the coefficients of, say, $j=2a-1$ and $j=2a$, for $a=1,2,3$. Moreover, the sign of the action on each coefficient is given by the action on the corresponding monomials in the basis of $S_x^5$ and $S_x^3$. Thus we find that the action of $\t$ is given by (say for $j=1$ and $j=2$):
\begin{multline}
\t: (a_{0,1}, a_{1,1}, a_{2,1}, a_{3,1}, a_{4,1}, a_{5,1}, b_{0,1}, b_{1,1}, b_{2,1}, b_{3,1} ) \\
\mapsto (a_{0,2}, - a_{1,2}, a_{2,2}, - a_{3,2}, a_{4,2}, - a_{5,2}, b_{0,2}, - b_{1,2}, b_{2,2}, - b_{3,2} ),
\end{multline}
and similarly for $j=3,4$ and $j=5,6$. The action on the last three points is a little bit more subtle, since in fact these three points are simply $\infty$. But it can be analyzed as in \cite{Donagi:2004ub}, and it turns out that there is a discrete choice involved, depending on whether $\t$ interchanges two of the rows or not. We will assume that $\t$ does not interchange the rows with $j=7,8,9$, and thus gives the action
\begin{multline}
\t: (a_{0,j}, a_{1,j}, a_{2,j}, a_{3,j}, a_{4,j}, a_{5,j}, b_{0,j}, b_{1,j}, b_{2,j}, b_{3,j} ) \\
\mapsto (a_{0,j}, - a_{1,j}, a_{2,j}, - a_{3,j}, a_{4,j}, - a_{5,j}, b_{0,j}, - b_{1,j}, b_{2,j}, - b_{3,j} ),
\end{multline}
for $j=7,8,9$.

We want an invariant extension class $[ \tV^* ]$. Thus we see that for $j=1,2$ (and similarly for $j=3,4$ and $j=5,6$), the coefficients must satisfy $a_{n,1} = a_{n,2}$ for $n$ even and $a_{n,1} = -a_{n,2}$ for $n$ odd, and similarly for the $b$ coefficients. For $j=7,8,9$, we simply have that $a_{n,j}$ vanishes for $n$ odd, and similarly for the $b$s.

Now we are ready to use these results to analyze the matrix $M$. The columns of $M$ have a definite parity under $\t$ given by the monomials in $S_x^{6*}$ and $S_x^{4*} \otimes S_t^1$ that they represent. Moreover, by combining the rows in a specific way we can also make the rows of $M$ have a definite parity under $\t$. Now, from the fact that the extension class is $\t$-invariant, which implies the above constraints on the coefficients $a_{i,j}$ and $b_{k,j}$, it is straigthforward to show that the matrix $M$ is block diagonal, with a $9 \times 9$ block taking $+$ to $+$ and a $9 \times 8$ block taking $-$ to $-$.

Furthermore, the $9 \times 8$ $-$ block breaks into a $6 \times 8$ subblock, that we call $A_-$, followed by a $3 \times 8$ subblock $B_-$. The $9 \times 9$ $+$ block also splits into a $6 \times 9$ subblock $A_+$ and a $3 \times 9$ subblock $B_+$. We see from \eqref{e:ma} and \eqref{e:mb} that the block $A_+$ is simply the block $A_-$ with a column of zeroes added to the right, while the block $B_+$ is the block $B_-$ with a column of zeroes added to the left. Schematically, $M$ is thus given by
\begin{equation}
\left(
\begin{array}{cccccc|cccccc}
&A_-&&0&&&&&\\
0 &&B_-&&&&&&\\
\hline
&&&&&&&~~~&A_-&~~~&\\
&&&&&&&~~~&B_-&~~~&\\
\end{array}
\right),
\end{equation}
where $0$ means a column of zeroes.

For generic coefficients in $A_-$ and $B_-$, and thus for a generic invariant extension class, the $9 \times 9$ $+$ block has rank $9$, and the $9 \times 8$ $-$ block has rank 8 (which means that there is a linear relation between the rows of $B_-$ and the rows of $A_-$). Thus $M$ has rank $17$. However, we can choose our bundle $\tV^*$ to be in a non-generic (but non-trivial) invariant extension class.\footnote{This simply corresponds to choosing a different bundle ${\tV^*}$ in the defining exact sequence \eqref{e:bdef}.} Then, it is easy to choose specific coefficients such that there are more than one linear relations between the rows of $B_-$ and the rows of $A_-$, which reduces the rank of the matrix. If there are linear relations inside $B_-$ or $A_-$, then both the $+$ block and the $-$ block have reduced rank; however, it is possible to choose linear relations between the rows of the $A_-$ and the rows of the $B_-$ such that only the rank of the $-$ block gets reduced. For example, take the rows of $A_-$ to be the unit vectors $e_1$ through $e_5$ and $e_7$, while $B_-$ has rows $e_5$, $e_7$, $e_8$. Then the $-$ block has rank $7$, while the $+$ block has rank $9$.

As we will see, phenomenologically we are interested in keeping the rank of the $+$ block maximal. In that case, the most we can do is to choose coefficients such that the three rows of $B_-$ are linearly dependent on the rows of $A_-$. Then the negative block has rank $6$, and the full matrix rank $15$. Any further reduction of the rank will also involve a reduction of the rank of the $9 \times 9$ $+$ block.

To summarize, we see that the rank of $M$ depends on the invariant extension class that we choose. Generically, $M$ has rank $17$, and splits into a $9$-dimensional $+$ subspace and a $8$-dimensional $-$ subspace. However, by choosing a non-generic invariant extension class, which we are allowed to do, we can reduce the rank of $M$. If we want to keep the $+$ subspace $9$-dimensional, this gives us two more choices; either the $-$ part has rank $7$ or rank $6$.

This translates into cohomology by the exact sequence \eqref{e:es}. As we saw, if the rank of $M$ is $17-n$ for $n$ integer, then we have $h^1(\tX, \wedge^2 \tV^*) = 6+n$ and $h^2(\tX, \wedge^2 \tV^*)=h^1(\tX, \wedge^2 \tV)=n $. From the above analysis, $n=0,1,2$. Moreover, we can compute the splitting of these cohomology groups into their invariant and anti-invariant parts. As in \cite{Donagi:2004ub}, we find that
\begin{align}
h^1 ( \tX, \wedge^2 \tV_2 )_+ &= 3, &h^1 ( \tX, \wedge^2 \tV_2 )_- &= 2,\notag\\
h^2 ( \tX, \wedge^2 \tV_2 )_+ &= 9, &h^2 ( \tX, \wedge^2 \tV_2 )_- &= 8,\notag\\
h^1 ( \tX, V_2 \otimes V_3)_+ &= 9, &h^1 ( \tX, V_2 \otimes V_3)_- &= 9.
\end{align}
Thus, we see that the relevant cohomology groups split as follows
\begin{align}
h^1(\tX, \wedge^2 \tV^*)_+ &= 3, &h^1(\tX, \wedge^2 \tV^*)_- &= 3+n,\notag\\
h^1(\tX, \wedge^2 \tV)_+ &= 0, &h^1(\tX, \wedge^2 \tV)_- &= n,
\end{align}
for $n=0,1,2$.

\section{The Particle Spectrum}\label{s:spectrum}

In this model, we compactify heterotic string theory on a Calabi-Yau threefold with an $SU(5)$ bundle; therefore the residual gauge group, which is the commutant of $SU(5)$ in $E_8$, is also $SU(5)$. Then, we use a $\IZ_2$ Wilson line (since our Calabi-Yau threefold has $\IZ_2$ fundamental group) to break the $SU(5)$ gauge group to the Standard Model gauge group $SU(3)_C \times SU(2)_L \times U(1)_Y$.

The particle spectrum of the compactification is given by the decomposition of the adjoint representation of $E_8$ under the above symmetry breaking pattern. More precisely, the multiplicity of the representations of the low energy gauge group are determined by the dimensions of the invariant and anti-invariant parts of the cohomology groups computed in the previous section. The details of the group action were explained in section 2 of \cite{Donagi:2004ub}.

\begin{table}[tb]
\begin{center}
\begin{tabular}{lcc}\hline
Multiplicity & Representation & Name\\
\hline\\[1pt]
$1 = h^0(\tX, \co_{\tX})_+$ & $(8,1)_0 \oplus (1,3)_0 \oplus (1,1)_0$ & Gauge connections of \\
&&$SU(3)_C$, $SU(2)_L$ and $U(1)_Y$\\
\hline\\[1pt]
$3 = h^1(\tX, \tV^*)_+$ & $({\bar 3},1)_{-4} \oplus (1,1)_6$ & left-handed anti-up and\\
&&left-handed charged anti-lepton\\
\hline\\[1pt] 
$3 = h^1(\tX, \tV^*)_-$ & $(3,2)_1$ & left-handed quark\\
\hline\\[1pt]
$0 = h^1(\tX, \tV)_+$ & $(3,1)_{4} \oplus (1,1)_{-6}$ & exotic\\
\hline\\[1pt]
$0 = h^1(\tX, \tV)_-$ & $({\bar 3},2)_{-1}$ & exotic\\
\hline\\[1pt]
$3 = h^1(\tX, \wedge^2 \tV^*)_+$ & $({\bar 3},1)_{2}$ & left-handed anti-down\\
\hline\\[1pt]
$3+n = h^1(\tX, \wedge^2 \tV^*)_-$ & $(1,{\bar 2})_{-3}$ & left-handed lepton and\\
&&down Higgs \\
\hline\\[1pt]
$0 = h^1(\tX, \wedge^2 \tV)_+$ & $({3},1)_{-2}$ & exotic\\
\hline\\[1pt]
$n = h^1(\tX, \wedge^2 \tV)_-$ & $(1,2)_{3}$ & up Higgs\\
\hline
\end{tabular}
\end{center}
\caption{The particle spectrum of the low-energy $SU(3)_C \times SU(2)_L \times U(1)_Y$ theory. Notice that all exotic particles come with $0$ multiplicity, and that the spectrum include $n$ copies of Higgs conjugate pairs, where $n=0,1,2$. In the table, the $U(1)$ charges listed are $w = 3Y$.}
\label{t:spectrum}
\end{table}

Using the results of the previous section, we can extract the full particle spectrum of the supersymmetric vacuum, which is presented in table \ref{t:spectrum}.\footnote{Only the left chiral list is presented, but the particles are accompanied by their CPT conjugate.} We obtain three families of quark and lepton superfields, one copy of the gauge connections, and no exotic matter. This is precisely the spectrum of the MSSM. Moreover, in our model we are free to choose between $0$, $1$ or $2$ Higgs doublet conjugate pairs, without adding exotic matter. This provides a new realization of the Higgs doublet-triplet splitting mechanism introduced by Wilson lines.

Additionally, the spectrum contains moduli. A lower bound on the number of moduli of our model goes as follows. The Calabi-Yau manifold $X$ has $11$ K\"ahler parameters and $11$ complex structure parameters. The spectral data of the bundle introduces $18=7+11$ moduli. Further, many more moduli come from the extension space. The $90$-dimensional space of extensions splits into a $50$-dimensional invariant subspace and a $40$-dimensional anti-invariant subspace. Since rescaling the extension does not change the bundle, we have a $49$-parameter family of invariant extensions. Generically, these yield $0$ Higgs pairs. The one Higgs region has codimension $2$: therefore we obtain $47$ invariant extension moduli. For this region, the number of moduli adds up to at least $87$ moduli.

Finally, recall that our model is in the strong coupling regime of the heterotic string. We obtained that there are $M5$-branes wrapping the effective curve described by $2 f \times \pt + 6 \pt \times f'$. However, we did not need to add a gauge instanton in the hidden sector to cancel the anomaly.

Of course, obtaining real low-energy particle physics requires more than just the right spectrum. We also have to compute the exact interaction terms in the effective Lagrangian, such as the Yukawa couplings, the Higgs $\m$-terms, etc. These terms can be computed from the cohomology groups, as explained for instance in \cite{Braun:2005xp}. We hope to report on that in the near future.

\section*{Appendix}

In this Appendix we show that $W_3$ can be chosen to be $\t$-invariant. Recall from section \ref{s:bundle} that the spectral data consists of an $a_B$-invariant smooth curve $C_3 \in |3e+3f|$ and a line bundle $N_3$ on $C_3$. We need to show that there exist line bundles $N_3 \in Pic^7 (C_3)$ such that $N_3={\bf T}_{C_3} (N_3)$. As explained in that section, this reduces to showing that the group $i_{C_3}^* Pic (B)$ (consisting of line bundles on $C_3$ which are restrictions of global bundles on $B$) contains a dense subset of $Pic^0 (C_3)$.

This in turn can be seen by degenerating $C_3$ to a certain singular, reducible curve $C_3' = {\bar C}_2 \cup f_{\infty} \cup e$. Here ${\bar C}_2$ is a generic curve in the linear system $|2e + 2f|$, and $f_{\infty}$ is the smooth elliptic fiber $f_{\infty} = \b^{-1} (\infty)$. We will show the density of $i_{C'_3}^* Pic (B)$ in $Pic^0 (C'_3)$; the density for a nearby smooth $C_3$ then follows from semicontinuity.

Note that:
\begin{align}
g({\bar C}_2) &=2, &g(f_{\infty}) &= 1, &g(e) = 0,\notag\\
f_{\infty} \cdot e &=1, &{\bar C}_2 \cdot e &= 0, &{\bar C}_2 \cdot f_{\infty} = 2.
\end{align}
It follows that the $e$ component is redundant: $Pic^0 (C'_3) = Pic^0 ({\bar C}_2 \cup f_{\infty})$. This $4$-dimensional group is a $\IC^*$-extension of the $3$-dimensional compact product group $A:= Pic^0 ({\bar C}_2) \times Pic^0 (f_{\infty})$. The required density amounts to showing that no proper closed subgroup of $Pic^0 (C'_3)$ can contain the image of $Pic(B)$.

But just as in lemma 4.7 of \cite{Donagi:2000sm}, the class of the $\IC^*$-extension is specified by a point of $A$, which determines an image point in $Pic^0 (f_{\infty})$. The latter can be varied continuously by moving ${\hat C}_2$ (and hence its two intersection points with $f_{\infty}$) within its linear system. In particular, the $\IC^*$-extension class for generic data is not a torsion point of $Pic^0 (f_{\infty})$, hence it is also not torsion in $A$. It follows that no proper subgroup of $Pic^0 (C'_3)$ can map onto $A$, so we are reduced to showing density of the image in $A$ itself.

Now density in the elliptic curve $Pic^0 (f_{\infty})$ follows since all that is needed is one point of infinite order. Density in $Pic^0 ({\bar C}_2)$ was proved in lemma 4.5 of \cite{Donagi:2000sm}. Finally, no proper subgroup of $A$ can map onto both $Pic^0 (f_{\infty})$ and $Pic^0 ({\bar C}_2)$: this can be seen by specializing ${\bar C}_2$, as in lemma 4.5 of \cite{Donagi:2000sm}, to a degenerate curve consisting of the section $e$ taken with multiplicity $2$, plus an arbitrary pair of $\b$-fibers which are interchanged by $\a_B$. In this limit, $Pic^0 ({\bar C}_2)$ becomes the product of these two fibers. Since these can be varied continuously, it follows that neither is isogenous to $f_{\infty}$. This completes the proof of the density of $Pic(B)$ in a generic $Pic^0 (C_3)$, and hence also of existence of $\t$-invariant bundles $W_3$.

\bibliographystyle{JHEP}

\clearpage

\bibliography{refs}

\end{document}